# Empirical Evaluation of Membership Inference Attacks on NLP Text Classifiers: A Baseline Study on SST-2

William Novak

Department of Math, Data & Technology

Minot State University

Minot, ND, 58707

william.novak@minotstateu.edu

Muhammad Abusaqer

Department of Math, Data & Technology

Minot State University

Minot, ND, 58707

muhammad.abusaqer@minotstateu.edu

Authors' version. Presented at the 58th Midwest Instruction and Computing Symposium (MICS 2026), Eau Claire, WI, March 27 to 28, 2026. Trustworthy Language Intelligence Lab, Minot State University.

## Abstract

Membership inference attacks (MIAs) try to determine whether a specific record was used to train a model, a privacy risk that matters in natural language processing (NLP), where training data can contain sensitive user text. This paper presents a controlled benchmark of membership inference vulnerability for text classification on the GLUE SST-2 sentiment dataset. A TF-IDF + Logistic Regression pipeline and a fine-tuned DistilBERT classifier are compared under a loss-threshold MIA, with utility measured by development accuracy and macro F1. DistilBERT reached 0.9466 accuracy and 0.9460 macro F1 against 0.8756 and 0.8727 for Logistic Regression, yet both models leaked membership signal (Attack AUC 0.5615 and 0.5800, respectively). Two mitigations were tested. Stronger regularization reduced leakage for Logistic Regression at a visible utility cost, whereas fine-tuning DistilBERT for 2 epochs instead of 3 reduced leakage with negligible accuracy loss. Lightweight training adjustments can improve the privacy-utility trade-off without complex defenses.

# 1. Introduction

Membership inference attacks (MIAs) have emerged as a significant privacy threat in machine learning because they aim to determine whether a specific record was used during model training [1]. Although this task may appear narrow, successful membership inference can reveal whether an individual's data was included in a sensitive dataset, which is particularly concerning in domains involving personal, medical, financial, or behavioral information [1], [3]. Prior work has shown that such leakage is often tied to model behavior rather than direct disclosure of training examples, which makes MIAs relevant even when only prediction outputs are exposed [1], [2].

Privacy risk is especially important in natural language processing (NLP). Text datasets frequently contain sensitive user-generated content, and modern NLP systems are increasingly deployed through prediction APIs and cloud-based services [3]–[5]. At the same time, much of the literature on text classification emphasizes predictive performance, while privacy evaluation is less frequently included as a first-class criterion. This imbalance creates a practical gap: a model may achieve strong utility while still retaining measurable membership-related signal. The connection between overfitting and privacy leakage has been well documented, suggesting that stronger task performance alone should not be interpreted as evidence of privacy safety [2].

This study addresses that gap through a controlled benchmark on the SST-2 sentiment classification task from GLUE [6], [7]. Two target models are compared: a traditional TF-IDF + Logistic Regression pipeline and a fine-tuned DistilBERT classifier [8]. This pairing is useful because it contrasts a lightweight, interpretable baseline with a compact transformer model under the same dataset, attack design, and evaluation protocol. Comparative benchmarking between traditional and transformer-based NLP models has shown that transformers often improve classification performance, but such comparisons typically focus on utility rather than privacy leakage [10]. The present study therefore extends that comparative perspective into the privacy domain.

The paper is guided by three research questions. First, how much membership information can be inferred from a traditional text classifier compared with a compact transformer on SST-2? Second, how closely is privacy leakage associated with overfitting-related behavior in this setting? Third, to what extent can simple mitigation choices reduce privacy leakage while preserving utility? To answer these questions, the study evaluates both models using a loss-threshold membership inference attack, reports both utility and privacy metrics, and tests two simple mitigations: stronger regularization for Logistic Regression and fewer fine-tuning epochs for DistilBERT. The resulting benchmark is intended to provide a clear empirical view of the privacy-utility trade-off in text classification.

# 2. Related Work

Membership inference was formalized by Shokri et al., who showed that an adversary can distinguish members from non-members by learning how a target model behaves on training versus unseen examples [1]. Their work demonstrated that even black-box access to prediction outputs may be sufficient to extract membership-related signal, and it established shadow-model-based attacks as a foundational approach in the area [1]. Soon after, Yeom et al. clarified the connection between membership inference and overfitting, showing that privacy leakage often increases when models generalize poorly and assign systematically lower loss to training points than to non-members [2]. These two studies remain central because they define both the attack setting and the intuition behind loss-based membership inference.

Subsequent work broadened the scope of MIA research. Niu et al. surveyed attack and defense mechanisms across supervised, generative, and deep-learning settings, emphasizing that privacy risk depends on model class, attack assumptions, and defense design [3]. Their survey highlights common defenses such as regularization, distillation, and differential privacy, while also noting that these defenses often introduce nontrivial performance trade-offs [3]. More recently, Wu and Cao extended the survey perspective to large-scale models, organizing MIAs by access scenario, including black-box, white-box, and intermediate settings, and showing that privacy threats remain relevant as model size and deployment scale increase [4]. Together, these surveys show that MIAs are no longer limited to narrow academic scenarios; they are now understood as a broader class of privacy risks across modern machine learning systems [3], [4].

Several studies have focused on aspects of membership inference beyond average attack success. Zhong et al. examined the disparate effects of MIAs and showed that privacy leakage may not be distributed uniformly across subpopulations [9]. Their results introduced the idea that privacy evaluation should consider fairness-related dimensions, not merely aggregate attack performance [9]. This perspective is important because it suggests that a model can appear moderately secure on average while still imposing uneven privacy risks across different groups.

NLP-specific membership inference has also received focused attention. Shejwalkar et al. studied user-level MIAs against NLP classification models and showed that privacy leakage can arise in text settings where data are grouped by user identity or contribution history [5]. Using Reddit and Amazon-based classification scenarios, they found that attack vulnerability depends partly on how much data are associated with each user and on how model behavior changes across data regimes [5]. Their work is especially relevant here because it directly confirms that NLP classifiers are viable targets for membership inference, while also showing that leakage patterns can be data-dependent rather than uniform [5].

Despite this growing literature, an important gap remains. Much of the prior work either focuses on broad surveys [3], [4], foundational theory and attack construction [1], [2], fairness and subgroup effects [9], or NLP-specific privacy exposure without direct comparison to traditional baselines [5]. By contrast, the present study provides a unified benchmark that compares a traditional sparse-feature classifier with a compact transformer on a standard NLP dataset under the same loss-threshold attack and the same utility/privacy evaluation framework. In that sense, the study is also methodologically aligned with comparative NLP benchmarking work showing the value of evaluating traditional and transformer-based models side by side [10], but it extends that logic to include privacy leakage rather than utility alone.

## 3. Methodology and Experimental Setup

This study evaluates privacy leakage in NLP text classifiers through a controlled benchmark of membership inference attacks (MIAs). The experiments compare a traditional text-classification pipeline, TF-IDF + Logistic Regression, with a compact transformer pipeline, DistilBERT, under the same dataset, attack design, and evaluation criteria. The goal is to examine both classification utility and privacy leakage, and to determine whether simple mitigation settings can reduce leakage with acceptable utility trade-offs [1], [2].

The experiments use the SST-2 sentiment classification task from the GLUE benchmark [6], [7]. SST-2 was selected because it is a standardized binary text-classification dataset [6], [7]. The official GLUE training split was used to build the target models, while the official validation split was reserved as a clean non-member pool for privacy evaluation. To support reproducibility, the official training split was partitioned using stratified sampling into target-train (80%) and target-dev (20%) subsets with a fixed random seed of 42.

Two target models were evaluated. The first was a TF-IDF + Logistic Regression baseline. Text was represented using lowercase normalization, English stopword removal, unigram and bigram features, min_df = 2, and max_features = 20,000. Logistic Regression was trained with the liblinear solver and max_iter = 1000. The baseline setting used C = 1.0, while the mitigation setting strengthened regularization by reducing the inverse regularization parameter to C = 0.3. The second target model was DistilBERT (distilbert-base-uncased), which was selected as a compact transformer alternative to full BERT [8]. Sentences were tokenized with truncation and padding to a maximum length of 128 tokens. Fine-tuning used batch size 16, learning rate $2 \times 10^{-5}$, and weight_decay = 0.01. The baseline DistilBERT configuration used 3 epochs, while the mitigation setting reduced training to 2 epochs.

Privacy evaluation was performed using a loss-threshold membership inference attack. In this setting, the attacker attempts to determine whether a specific example was used during model training [1]. Following prior work, the attack uses per-example negative log-likelihood as its primary signal, since training-set members often receive lower loss than non-members when models exhibit overfitting-related behavior [1], [2]. For each example $(x, y)$, the loss is computed as

$$-\log p_\theta(y \mid x)$$

where $p_\theta(y \mid x)$ denotes the predicted probability assigned to the true class. An attack score is then defined as the negative of this loss so that higher scores indicate a greater likelihood of membership. For privacy evaluation, a balanced attack set was constructed by sampling an equal number of member examples from the target-train subset and non-member examples from the official validation split.

The study reports both utility metrics and privacy metrics. Utility was measured using development accuracy and macro F1-score. In addition, the mean per-example loss was recorded for the target-train, target-dev, and validation splits to help characterize model fit and generalization. Privacy leakage was measured using Attack AUC and Attack Advantage. Attack AUC quantifies how well the attack score separates members from non-members across thresholds, while Attack Advantage is defined as

$$\max_{\tau}(\mathrm{TPR}(\tau) - \mathrm{FPR}(\tau))$$

where $\tau$ is the decision threshold. Attack accuracy at the best threshold was also reported as a supplementary measure. These metrics enable direct examination of the privacy-utility trade-off rather than classification performance alone [2], [10].

All experiments were implemented in Python using a Jupyter Notebook workflow. The traditional pipeline used scikit-learn for TF-IDF feature extraction and Logistic Regression, while the transformer pipeline used Hugging Face Transformers and PyTorch for DistilBERT fine-tuning and inference. Intermediate outputs, including predicted probabilities, per-example losses, attack summaries, and comparison tables, were saved to support reproducibility and streamline subsequent analysis. This comparative workflow is consistent with prior benchmarking studies that contrast traditional and transformer-based text-classification models under shared evaluation settings [10].

# 4. Results

This section reports the baseline comparison between TF-IDF + Logistic Regression and DistilBERT, followed by the effect of the two mitigation settings evaluated in this study. Utility is reported using development accuracy and macro F1-score, whereas privacy leakage is reported using Attack AUC and Attack Advantage. Mean train, development, and validation losses are also examined to contextualize the observed privacy-utility trade-offs.

## 4.1. Baseline Model Comparison

As shown in Table 1, DistilBERT achieved substantially higher predictive utility than the TF-IDF + Logistic Regression baseline. DistilBERT reached a development accuracy of 0.9466 and a macro F1-score of 0.9460, compared with 0.8756 and 0.8727 for Logistic Regression. Loss statistics also favored DistilBERT, particularly on the training and development splits.

| Model | Dev Acc. | Macro F1 | Train Loss | Dev Loss | Val. Loss | Attack AUC | Attack Adv. | Attack Acc. |
|---|---|---|---|---|---|---|---|---|
| TF-IDF + LR | 0.8756 | 0.8727 | 0.3105 | 0.3501 | 0.4553 | 0.5615 | 0.1697 | 0.5849 |
| DistilBERT | 0.9466 | 0.9460 | 0.0510 | 0.2341 | 0.4471 | 0.5800 | 0.1365 | 0.5682 |

Table 1: Baseline comparison of model utility and privacy metrics

From a privacy perspective, however, both models remained vulnerable to the loss-threshold membership inference attack. DistilBERT produced the higher Attack AUC (0.5800 versus 0.5615), indicating slightly stronger average separability between members and non-members, while Logistic Regression produced the higher Attack Advantage (0.1697 versus 0.1365), indicating stronger threshold-specific separation at its best operating point. Figures 1 and 2 visualize these baseline differences.

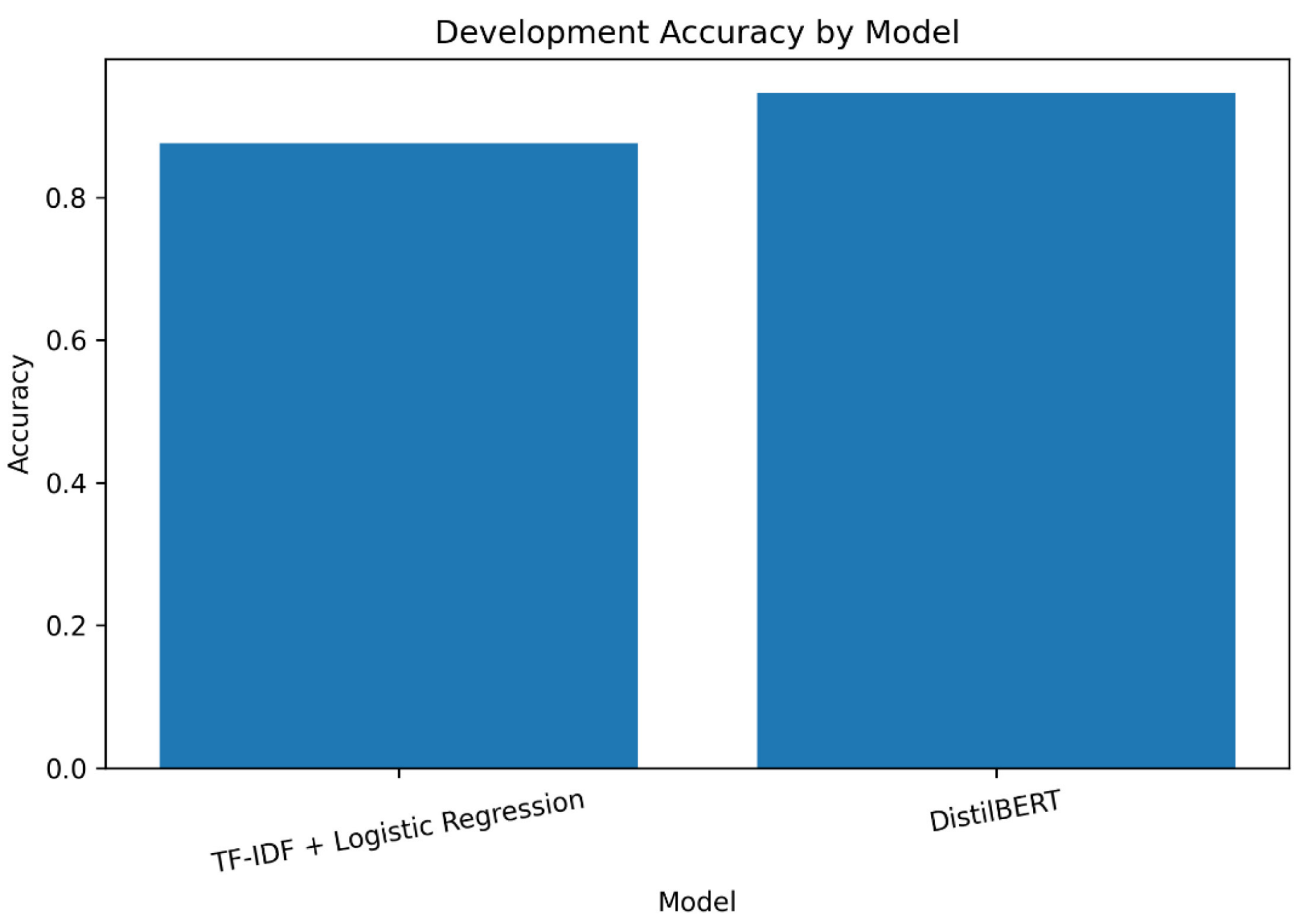

Figure 1: Development accuracy for the two baseline models.

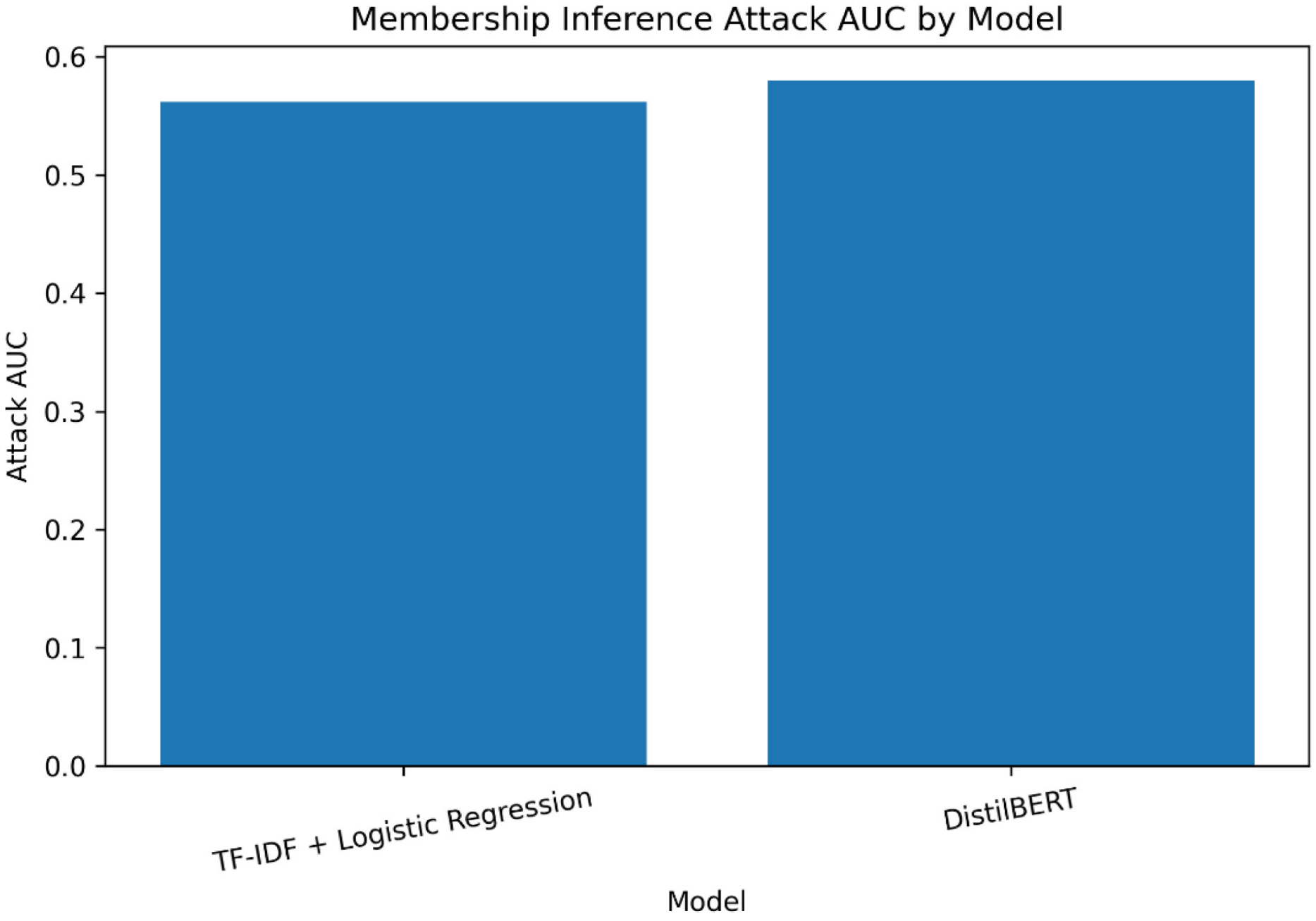

Figure 2: Attack AUC for the two baseline models.

## 4.2. Effect of Mitigation Settings

Table 2 and Figures 3 and 4 summarize the mitigation runs. For Logistic Regression, increasing regularization from C = 1.0 to C = 0.3 reduced Attack AUC from 0.5615 to 0.5385 and Attack Advantage from 0.1697 to 0.1330. These privacy improvements were accompanied by a noticeable utility cost: development accuracy decreased by 0.0187, and macro F1 decreased by 0.0202. Mean losses increased across the train, development, and validation splits (0.3105→0.4151, 0.3501→0.4358, and 0.4553→0.4894), indicating a less confident model.

For DistilBERT, reducing training from 3 to 2 epochs also lowered privacy leakage, decreasing Attack AUC from 0.5800 to 0.5654 and Attack Advantage from 0.1365 to 0.1170. In contrast to Logistic Regression, the utility cost was negligible: development accuracy changed from 0.9466 to 0.9465, and macro F1 changed from 0.9460 to 0.9458. Moreover, while train mean loss increased (0.0510→0.0771), both development and validation losses decreased (0.2341→0.2005 and 0.4471→0.3777), which is consistent with reduced overfitting under the 2-epoch setting.

| **Model family** | **Δ Dev Acc.** | **Δ Macro F1** | **Δ Attack AUC** | **Δ Attack Adv.** |
|---|---|---|---|---|
| Logistic Regression | -0.0187 | -0.0202 | -0.0230 | -0.0367 |
| DistilBERT | -0.0001 | -0.0002 | -0.0145 | -0.0195 |

Table 2: Change from baseline to mitigation setting within each model family

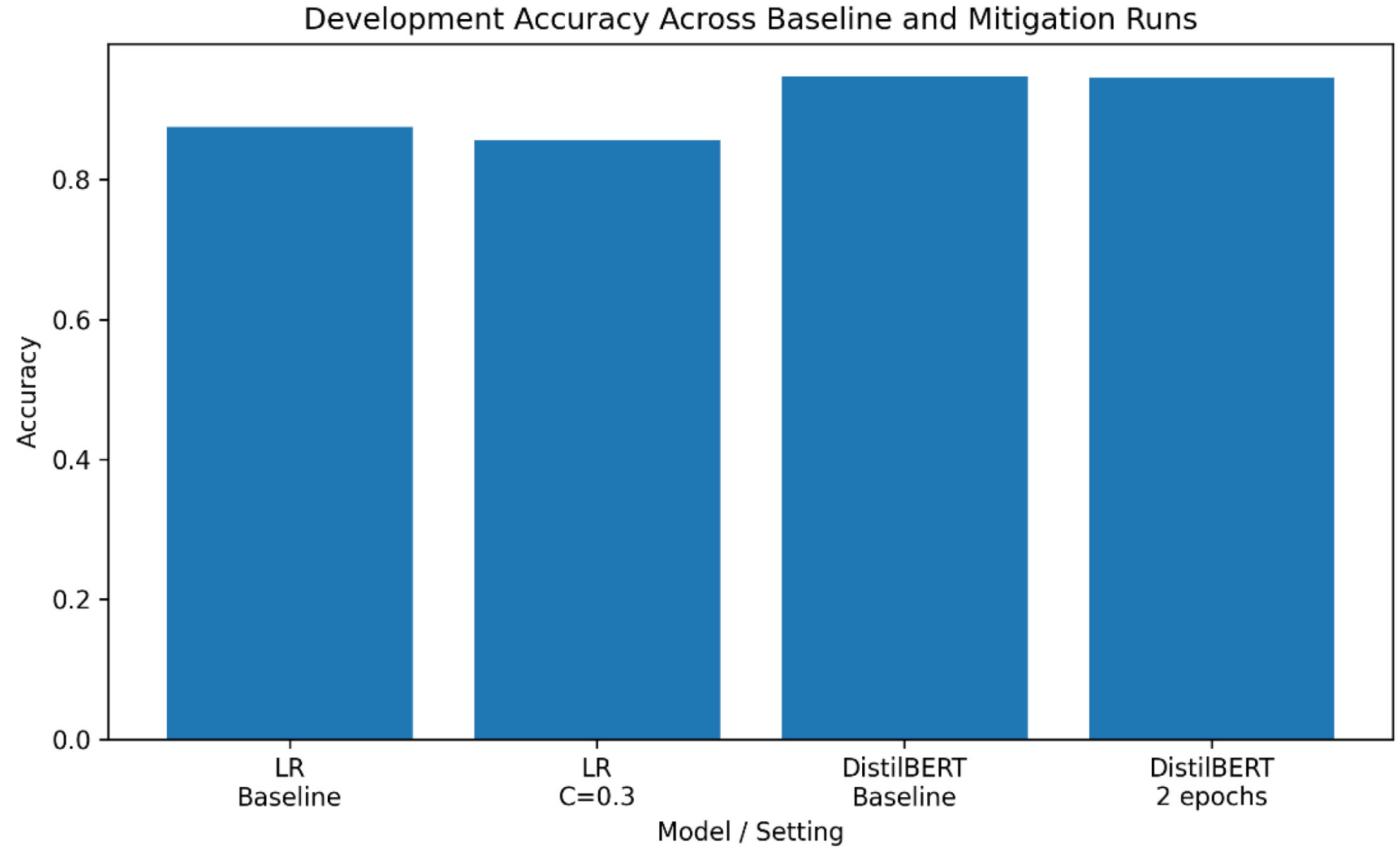

Figure 3: Development accuracy across baseline and mitigation runs.

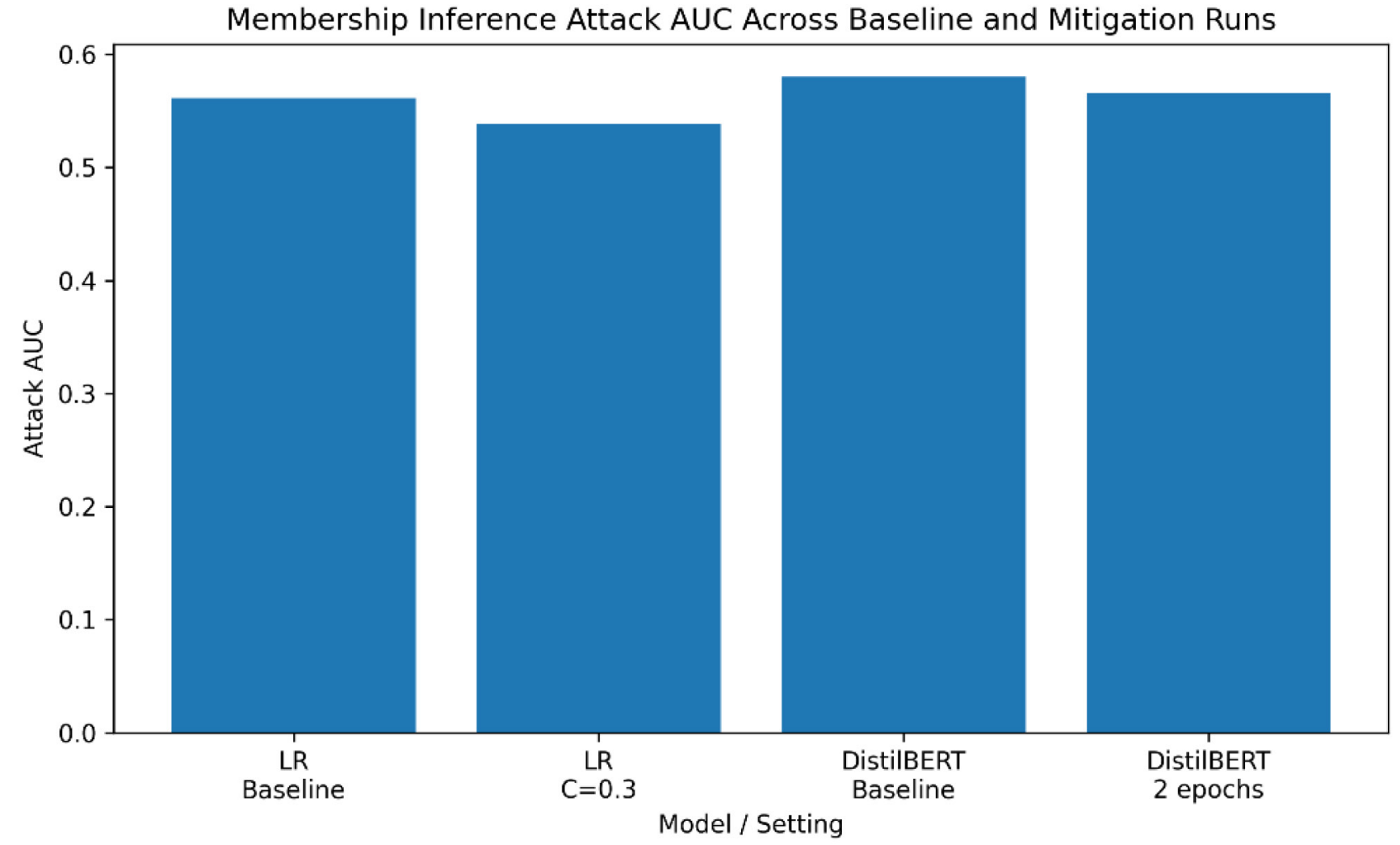

Figure 4: Attack AUC across baseline and mitigation runs.

## 4.3. Summary of Observed Trade-offs

Across both model families, the evaluated mitigation settings reduced attack effectiveness. Stronger regularization improved privacy metrics for Logistic Regression but lowered utility, whereas training DistilBERT for 2 epochs reduced leakage with almost no utility loss. Overall, DistilBERT with 2 epochs yielded the most favorable privacy-utility trade-off among the evaluated settings.

## 5. Discussion

The results show a consistent privacy-utility pattern across the two model families. On the utility side, DistilBERT produced clearly stronger classification performance than the TF-IDF + Logistic Regression baseline, improving development accuracy from 0.8756 to 0.9466 and macro F1 from 0.8727 to 0.9460. This outcome is consistent with prior findings that transformer-based models often outperform traditional linear baselines on sentence-level NLP tasks because they capture richer contextual information than sparse lexical representations [8], [10]. At the same time, both target models exhibited measurable membership leakage under the loss-threshold attack, with Attack AUC values above random guessing (0.5615 for Logistic Regression and 0.5800 for DistilBERT). These values indicate that the learned models retained enough membership-related signal to permit non-trivial separation between members and non-members, which is aligned with the general threat model described in prior membership inference work [1], [2], [3].

### 5.1. Baseline Comparison

A more detailed comparison reveals that utility and privacy did not move in exactly the same way across metrics. DistilBERT achieved the best utility and also the highest Attack AUC, suggesting that its outputs were, on average, slightly more rank-separable for the attack than those of Logistic Regression. However, Logistic Regression yielded the higher Attack Advantage (0.1697 versus 0.1365). This difference is important because Attack AUC is a threshold-independent ranking measure, while Attack Advantage reflects the best separation achievable at a specific threshold. In practical terms, the results suggest that DistilBERT exposed somewhat stronger global membership signal, whereas Logistic Regression produced a threshold region with a larger local separation between true positive rate and false positive rate. This distinction shows that privacy risk should not be interpreted from a single metric alone, especially in small and controlled experiments [1], [3].

The loss statistics also support the privacy interpretation. For Logistic Regression, the mean loss increased from 0.3105 on target-train data to 0.3501 on target-dev and 0.4553 on the validation split. For DistilBERT, the corresponding values were 0.0510, 0.2341, and 0.4471. In both cases, member examples received lower average loss than non-members, which is exactly the signal exploited by the loss-threshold attack. This behavior matches the established connection between overfitting-related generalization gaps and membership inference vulnerability [2]. The transformer model showed much lower training loss than the traditional baseline, indicating a better fit to the training set, but it did not produce a dramatically larger privacy risk. Instead, the observed leakage remained moderate, suggesting that higher predictive performance alone does not automatically imply severe privacy exposure in this benchmark setting.

## 5.2. Effect of the Mitigation Settings

The mitigation experiments provide the most practically useful result of the study. For Logistic Regression, stronger regularization (C = 0.3) reduced Attack AUC from 0.5615 to 0.5385 and reduced Attack Advantage from 0.1697 to 0.1330. However, this improvement in privacy came with a noticeable utility cost: development accuracy dropped by 0.0187 and macro F1 dropped by 0.0202. This pattern matches the usual expectation that stronger regularization can reduce memorization-related leakage, but may also suppress useful task signal if the model becomes too constrained [2], [3]. In other words, the Logistic Regression mitigation succeeded in reducing leakage, but its privacy gains were purchased with a meaningful degradation in classification performance.

The DistilBERT mitigation produced a more favorable trade-off. Reducing the number of fine-tuning epochs from 3 to 2 lowered Attack AUC from 0.5800 to 0.5654 and reduced Attack Advantage from 0.1365 to 0.1170, while development accuracy remained essentially unchanged (0.9466 versus 0.9465). Macro F1 also remained nearly identical. This result is particularly important because it suggests that a small training adjustment can reduce privacy leakage without sacrificing model utility in any meaningful way. The loss values further support this interpretation: under the 2-epoch setting, the mean development and validation losses both decreased relative to the baseline, while the mean training loss increased from 0.0510 to 0.0771. This pattern is consistent with improved generalization and reduced overfitting pressure, which in turn reduces the effectiveness of membership inference [2].

## 5.3. Practical Interpretation

From a practical cybersecurity perspective, the results show that privacy leakage in NLP classifiers can be measurable even in a compact, small-scale benchmark, but the leakage observed here was modest rather than extreme. All Attack AUC values remained close to the 0.54 to 0.58 range, indicating that the attack was stronger than random guessing but still far from highly reliable extraction of membership status. This is an important outcome because it supports a balanced interpretation: privacy risk is real and quantifiable, yet it should not be overstated when the observed signal is only moderate.

The experiments also suggest that the choice of mitigation should depend on the model family. For the traditional pipeline, privacy reduction was achieved primarily through stronger regularization, but with a visible utility penalty. For the transformer pipeline, earlier stopping preserved nearly all utility while still reducing the attack metrics. As a result, the transformer mitigation produced the strongest privacy-utility trade-off among the settings evaluated in this study. This finding is consistent with the broader idea that privacy-aware model selection should focus not only on attack resistance, but also on whether a mitigation remains operationally acceptable for the target task [1], [2], [3].

### 5.4. Scope of the Findings

These conclusions should be interpreted within the scope of the current benchmark. The study used a single dataset, a single attack family, and a small set of mitigation settings. Therefore, the results should be understood as a compact empirical baseline rather than a universal claim about privacy leakage in all NLP systems. Nevertheless, the trends are meaningful. The baseline experiments confirmed measurable membership signal in both model types, and the mitigation runs showed that simple design choices can reduce that signal.

## 6. Future Work

Several extensions would strengthen this line of work. First, future experiments could evaluate additional datasets beyond SST-2 to determine whether the trends observed here remain stable across different text domains, class distributions, and levels of task difficulty. Second, the attack space could be expanded beyond the current loss-threshold design to include stronger shadow-model-based or confidence-based attacks, which would provide a broader view of privacy risk under alternative adversarial assumptions [1], [3]. Third, the mitigation space could be extended with stronger weight decay, calibration-based approaches, or formal privacy-preserving methods such as differential privacy training. Such additions would allow future studies to compare simple heuristic mitigations against more principled defenses.

From a project-development perspective, the most natural next step is to broaden the experimental matrix while preserving the same reproducible workflow established in this paper. A larger follow-up study could compare additional transformer settings, more regularization choices, and multiple privacy metrics in a unified benchmark. This would help determine whether the favorable behavior observed for the fewer-epochs DistilBERT mitigation generalizes beyond the current setup. More broadly, future work may also investigate whether similar privacy-utility trade-offs arise in cybersecurity-oriented NLP tasks, where textual data can be more sensitive and privacy risks may be operationally significant [3], [5].

## 7. Conclusion

This study presented a compact benchmark of membership inference risk for NLP text classifiers using the SST-2 dataset and a loss-threshold attack setting. Two target models were evaluated: a traditional TF-IDF + Logistic Regression pipeline and a DistilBERT-based transformer pipeline. The experiments showed that DistilBERT achieved the strongest task performance, reaching a development accuracy of 0.9466 and a macro F1-score of 0.9460, compared with 0.8756 and 0.8727, respectively, for the Logistic Regression baseline. At the same time, both models exhibited measurable membership leakage, with Attack AUC values above random guessing (0.5615 for Logistic Regression and 0.5800 for DistilBERT), confirming that non-trivial membership signal remained present in both model families.

The mitigation experiments further showed that simple design choices can reduce privacy leakage, although the privacy-utility trade-off differed across models. For Logistic Regression, stronger regularization (C = 0.3) lowered Attack AUC from 0.5615 to 0.5385 and reduced Attack Advantage from 0.1697 to 0.1330, but it also decreased development accuracy by 0.0187. In contrast, reducing DistilBERT training from 3 epochs to 2 lowered Attack AUC from 0.5800 to 0.5654 and reduced Attack Advantage from 0.1365 to 0.1170 while leaving development accuracy essentially unchanged (0.9466 versus 0.9465). Among the mitigation settings tested, the fewer-epochs DistilBERT configuration produced the most favorable privacy-utility trade-off. These findings are consistent with prior work linking overfitting-related behavior to membership inference vulnerability and with comparative benchmarking studies showing that model quality should be assessed together with privacy risk rather than in isolation [2], [3], [10].

Overall, the results support three main conclusions. First, stronger NLP utility does not eliminate privacy risk. Second, privacy leakage in this benchmark was measurable but moderate. Third, lightweight mitigation choices can reduce leakage without necessarily imposing severe accuracy loss. Taken together, the experiments provide a clear, reproducible, and pedagogically accessible baseline for understanding membership inference behavior in text classifiers. This benchmark demonstrates the core privacy-utility relationship that motivates broader research on privacy-preserving machine learning [1], [2], [3].